\documentclass[twocolumn,a4paper]{revtex4}
\usepackage{amsfonts,amsmath,amssymb,mathrsfs,dsfont,yfonts,bbm}
\usepackage[shortlabels]{enumitem}
\usepackage[normalem]{ulem}
\usepackage{tikz}
\begin{document}

\newcommand{\Supertwistor}{\Cset \mathrm{P}^{3|4}}
\newcommand{\Twistorspace}{\Cset \mathrm{P}^{3}}
\newcommand{\half}{\frac{1}{2}}
\newcommand{\diff}{\mathrm{d}}
\newcommand{\ra}{\rightarrow}
\newcommand{\Zset}{{\mathbb Z}}
\newcommand{\Cset}{{\,\,{{{^{_{\pmb{\mid}}}}\kern-.47em{\mathrm C}}}}}
\newcommand{\Rset}{{\mathrm{I}\!\mathrm{R}}}
\newcommand{\gra}{\alpha}
\newcommand{\grl}{\lambda}
\newcommand{\gre}{\epsilon}
\newcommand{\zb}{{\bar{z}}}
\newcommand{\mn}{{\mu\nu}}
\newcommand{\Acal}{{\mathcal A}}
\newcommand{\Rcal}{{\mathcal R}}
\newcommand{\Dcal}{{\mathcal D}}
\newcommand{\Mcal}{{\mathcal M}}
\newcommand{\Ncal}{{\mathcal N}}
\newcommand{\Kcal}{{\mathcal K}}
\newcommand{\Lcal}{{\mathcal L}}
\newcommand{\Scal}{{\mathcal S}}
\newcommand{\CW}{{\mathcal W}}
\newcommand{\Bcal}{\mathcal{B}}
\newcommand{\Ccal}{\mathcal{C}}
\newcommand{\Vcal}{\mathcal{V}}
\newcommand{\Ocal}{\mathcal{O}}
\newcommand{\Zcal}{\mathcal{Z}}
\newcommand{\Zb}{\overline{Z}}
\newcommand{\Urm}{{\mathrm U}}
\newcommand{\Srm}{{\mathrm S}}
\newcommand{\SO}{\mathrm{SO}}
\newcommand{\Sp}{\mathrm{Sp}}
\newcommand{\SU}{\mathrm{SU}}
\newcommand{\U}{\mathrm{U}}
\newcommand{\be}{\begin{equation}}
\newcommand{\ee}{\end{equation}}
\newcommand{\Comment}[1]{{}}
\newcommand{\tQ}{\tilde{Q}}
\newcommand{\tq}{{\tilde{q}}}
\newcommand{\trho}{\tilde{\rho}}
\newcommand{\tphi}{\tilde{\phi}}
\newcommand{\Qcal}{\mathcal{Q}}
\newcommand{\tmu}{\tilde{\mu}}
\newcommand{\dbar}{\bar{\partial}}
\newcommand{\p}{\partial}
\newcommand{\eg}{{\it e.g.\;}}
\newcommand{\ie}{{\it i.e.\;}}
\newcommand{\Tr}{\mathrm{Tr}}
\newcommand{\twistor}{\Cset \mathrm{P}^{3}}
\newcommand{\note}[2]{{\footnotesize [{\sc #1}}---{\footnotesize   #2]}}
\newcommand{\CL}{\mathcal{L}}
\newcommand{\CJ}{\mathcal{J}}
\newcommand{\CA}{\mathcal{A}}
\newcommand{\CH}{\mathcal{H}}
\newcommand{\CD}{\mathcal{D}}
\newcommand{\CE}{\mathcal{E}}
\newcommand{\CQ}{\mathcal{Q}}
\newcommand{\CB}{\mathcal{B}}
\newcommand{\CC}{\mathcal{C}}
\newcommand{\CO}{\mathcal{O}}
\newcommand{\CT}{\mathcal{T}}
\newcommand{\CI}{\mathcal{I}}
\newcommand{\CN}{\mathcal{N}}
\newcommand{\CS}{\mathcal{S}}
\newcommand{\CM}{\mathcal{M}}

\parskip 11pt
\title{\Large {\bf $\CN=4$ SYM, Argyres-Douglas Theories, and an Exact~Graded~Vector~Space~Isomorphism}} 
\author {Matthew Buican$^{\dagger}$ and Takahiro Nishinaka$^{*}$} 
\affiliation{$^{\dagger}$CRST and School of Physics and Astronomy \\ Queen Mary University of London, London E1 4NS, UK\\ $^{*}$Department of Physical Sciences, College of Science and Engineering\\ Ritsumeikan University, Shiga 525-8577, Japan\\}

\begin{abstract}
\noindent
In this first of two papers, we explain in detail the simplest example of a broader set of relations between apparently very different theories. Our example relates $\mathfrak{su}(2)$ $\CN=4$ super Yang-Mills (SYM) to a theory we call \lq\lq$(3,2)$." This latter theory is an exactly marginal diagonal $SU(2)$ gauging of three $D_3(SU(2))$ Argyres-Douglas (AD) theories. We begin by observing that the Schur indices of these two theories are related by an algebraic transformation that is surprisingly reminiscent of index transformations describing spontaneous symmetry breaking on the Higgs branch. However, this transformation breaks half the supersymmetry of the SYM theory as well as its full $\CN=2$ $SU(2)_F$ flavor symmetry. Moreover, it does so in an interesting way when viewed through the lens of the corresponding 2D vertex operator algebras (VOAs): affine currents of the small $\CN=4$ super-Virasoro algebra at $c=-9$ get mapped to the $\CA(6)$ stress tensor and some of its conformal descendants, while the extra supersymmetry currents on the $\CN=4$ side get mapped to higher-dimensional fermionic currents and their descendants on the $\CA(6)$ side. We prove these relations are facets of an exact graded vector space isomorphism (GVSI) between these two VOAs. This GVSI respects the $U(1)_r$ charge of the parent 4D theories. We briefly sketch how more general $\mathfrak{su}(N)$ $\CN=4$ SYM theories are related to an infinite class of AD theories via generalizations of our example. We conclude by showing that, in this class of theories, the $\CA(6)$ VOA saturates a new inequality on the number of strong generators.
\end{abstract}
\maketitle

\section*{Introduction}
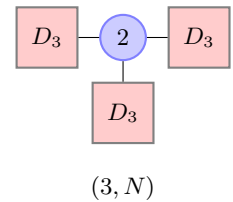
\begin{figure}
 \begin{center}
\begin{tikzpicture}[place/.style={circle,draw=blue!50,fill=blue!20,thick,inner sep=0pt,minimum size=6mm},transition2/.style={rectangle,draw=black!50,fill=red!20,thick,inner sep=0pt,minimum size=8mm},auto]

 \node[place] (1) at (0,0) [shape=circle] {\;$2$\;};
 \node[transition2] (2) at (-1,0) {\;$D_3$\;} edge (1);
 \node[transition2] (3) at (1,0) {\;$D_3$\;} edge (1);
 \node[transition2] (4) at (0,-1) {\;$D_3$\;} edge (1); 
 \node at (0,-2) {$(3,N)$}; 
 
  \end{tikzpicture}
\caption{Our AD theory of interest, $(3,2)$, consists of an exactly marginal diagonal $SU(2)$ gauging of three $D_3(SU(2))$ SCFTs.}
\label{fig:32}
 \end{center}
\end{figure}

In this note, we revisit a 4D $\CN=2$ superconformal field theory (SCFT) we first studied in \cite{Buican:2016arp} and find some remarkable relations it has to $\mathfrak{su}(2)$ $\CN=4$ super Yang-Mills (SYM). We will refer to the $\CN=2$ theory in question as the $(3,2)$ SCFT \footnote{In \cite{Buican:2016arp}, we referred to this SCFT somewhat unimaginatively as $\hat{\CT}$. It has also been studied in \cite{DelZotto:2015rca}, where it was called the $(E_6^{1,1},SU(2))$ SCFT.}. This theory consists of three copies of the so-called $D_3(SU(2))\simeq (A_1, A_3)$ Argyres-Douglas (AD) theory \footnote{These theories were originally discovered in \cite{Argyres:1995xn}. The nomenclature is borrowed from \cite{Cecotti:2012jx,Cecotti:2013lda,Cecotti:2010fi}.} with a gauged diagonal $SU(2)$ symmetry having vanishing beta function (see figure \ref{fig:32}). In what follows, we will refer to these $D_3(SU(2))$ SCFTs as \lq\lq$D_3$" theories.

Modulo having the same $\mathfrak{su}(2)$ gauge algebra, $\mathfrak{su}(2)$ $\CN=4$ SYM and the $(3,2)$ theory seem to be very different beasts. For example, in the former case, the $\CN=2$ matter sector (prior to gauging) consists of a free adjoint-valued hypermultiplet while, in the latter case, it consists of three strongly interacting $D_3$ SCFTs.

However, this picture begins to change when one thinks of the so-called \lq\lq Schur sector" of operators in both SCFT matter sectors (prior to gauging). In particular, the resemblance becomes starker when one thinks in terms of the related 2D vertex operator algebras (VOAs) that the general correspondence in \cite{Beem:2013sza} assigns to the matter sectors in question. On the SYM side, we have a VOA (strongly) generated by adjoint-valued dimension $1/2$ symplectic bosons while, on the $(3,2)$ side, we have a VOA (strongly) generated by adjoint-valued affine currents for each $D_3$ matter sector \cite{Buican:2015ina,Buican:2015hsa,Cordova:2015nma,Buican:2015tda}.

This realization motivates a cursory glance at some of the basic observables of the 4D theories we are discussing and reveals the following: in both cases $a=c$, and, as we will explain in section \ref{indexrel}, the Schur indices of the SYM theory and $(3,2)$ are related in a simple way
\begin{equation}\label{indrel}
\CI_{(3,2)}(q)=\CI_{\mathfrak{su}(2)}(q^3;q^{1\over2})~,
\end{equation}
where the label \lq\lq$\mathfrak{su}(2)$" refers to the corresponding $\CN=4$ theory. In writing \eqref{indrel}, we have set $x=q^{1\over2}$, where $x$ is the flavor $SU(2)_F$ fugacity that arises when we think of the $\CN=4$ theory as an $\CN=2$ theory.

As we will explain in section \ref{GVSI}, these features have a deeper
explanation in terms of a mathematically precise exact graded vector
space isomorphism (GVSI) between the 2D vertex operator algebras (VOAs)
that correspond to $(3,2)$ \cite{Buican:2016arp} and $\mathfrak{su}(2)$
$\CN=4$ SYM \cite{Beem:2013sza}. In other words, we find a GVSI between
the $\CA(6)$ algebra of \cite{Feigin:2007sp,Feigin:2008sg} and the small $\CN=4$ super-Virasoro algebra at $c=-9$.

We suggest that the GVSI and its consequences can be thought of physically as comprising a distant cousin of spontaneous symmetry breaking on the Higgs branch that we call \lq\lq distorted symmetry breaking." At the level of moduli spaces there is a superficial similarity to moving onto the Higgs branch and removing a decoupled Nambu-Goldstone (NG) multiplet: $\mathfrak{su}(2)$ $\CN=4$ SYM (thought of as an $\CN=2$ theory) has a 2-complex dimensional Higgs branch, while the $(3,2)$ theory has a trivial Higgs branch. Moreover, just as in the case of two theories related by Higgs branch renormalization group flows, the $(3,2)$ and $\mathfrak{su}(2)$ $\CN=4$ SYM theories have the same value of the $U(1)_r\subset SU(2)_R\times U(1)_r$ anomaly, namely
\begin{equation}
{\rm Tr}|_{(3,2)}\ U(1)_r={\rm Tr}|_{\mathfrak{su}(2)}\ U(1)_r=0~.
\end{equation}
This result follows from the fact that $a_{(3,2)}-c_{(3,2)}=a_{\mathfrak{su}(2)}-c_{\mathfrak{su}(2)}$.

Yet another similarity arises when one thinks of moving onto the Higgs branch in terms of the index \cite{Gaiotto:2012xa}. More precisely, to describe the Higgsing of the $\CN=4$ theory in terms of the index, we can, just as in \eqref{indrel}, set $x=q^{1\over2}$. Performing this substitution corresponds to leaving unsuppressed certain contributions to the Schur index from the $SU(2)_F$ lowest-weight component of the holomorphic moment map, $\mu_-$. The resulting divergence of the index is interpreted as setting $\langle\mu_-\rangle\ne0$ and moving onto the moduli space. The corresponding residue of the index describes the IR theory that remains (after removing the NG multiplet).

From the perspective of the index, only the fact that we simultaneously rescale $q\to q^3$ in \eqref{indrel} keeps $\CI_{\mathfrak{su}(2)}$ finite and tells the index to describe physics different from Higgsing. As we will see, this rescaling leads to interesting phenomena like the fact that, at the level of the corresponding VOAs, some of the (strong) generators of the $c=-9$ small $\CN=4$ super-Virasoro algebra are mapped to generators of the $\CA(6)$ chiral algebra while others are mapped to descendants. More generally, the number of derivatives acting on an operator in one theory is not preserved under the mapping to its cousin in the other theory (see section \ref{GVSI} for details).

However, even this peculiar mapping of operators is somewhat reminiscent of spontaneous symmetry breaking. Indeed, consider a flavor symmetry current for some spontaneously broken symmetry in a general quantum field theory (QFT) in $d>2$. In the deep IR, this current is mapped to a descendant of the NG field 
\begin{equation}\label{pion}
j_{\mu}\to f_{\pi}\partial_{\mu}\pi~.
\end{equation}
More generally, we may expect the following mapping between operators in the UV and IR
\begin{equation}\label{UVtoIR}
\partial^n\CO_{UV}\to\partial^m\CO_{IR}~,
\end{equation}
where it may happen that $n\ne m$ \footnote{We have been schematic about contraction of indices above, but both sides of the above relation must transform in the same way under the Lorentz group.}.

In the context of $\CN=2$ theories, it is useful to think of the above discussion in terms of Higgs branch operators and their Hall-Littlewood (HL) generalizations \cite{Beem:2013sza,Gadde:2011uv}. Thinking along these lines, we easily find examples of \eqref{pion} and \eqref{UVtoIR}. Such situations often arise due to the fact that certain Higgs branch operators of the UV theory get mapped to operators that, in the deep IR, have support only in the low energy effective QFT describing the decoupled NG multiplets and vanish in the remaining IR QFT (e.g., the $\CN=2$ flavor symmetry current multiplets for spontaneously broken flavor symmetries). In terms of symbols, we have 
\begin{equation}\label{flowtozero}
\CO_{\rm UV}^{\rm HL}\to\CO_{\rm IR}|_{{\rm rem.}}=0~,
\end{equation}
where $\CO_{UV}^{\rm HL}$ is an HL operator in the UV theory, $\CO_{IR}\ne0$ is its IR avatar supported in the NG effective QFT, and \lq\lq$|_{\rm rem.}$" denotes the restriction of this operator to the part of the IR theory decoupled from the NG multiplets.

We will see that many of these phenomena have cousins in the case of distorted symmetry breaking. For example, all HL operators of the $\mathfrak{su}(2)$ $\CN=4$ SYM theory will be mapped to non-HL operators in the $(3,2)$ theory (whose Higgs branch and HL sector vanishes). In other words, instead of \eqref{flowtozero}, we will have
\begin{equation}\label{distorted}
\CO_{\mathfrak{su}(2)}^{\rm HL}\to\CO_{(3,2)}|_{\rm HL}=0~,
\end{equation}
where $\CO_{(3,2)}\ne0$ is an operator of the $(3,2)$ SCFT, and \lq\lq$|_{\rm HL}$" denotes the restriction to the (trivial) HL ring.

The plan of this paper is as follows. In the next section we describe the ingredients that lead to the index relation \eqref{indrel}. With this groundwork out of the way, in section \ref{GVSI} we proceed to describe the GVSI between the VOAs discussed above. In section \ref{ineq}, we tease a generalization of the relation between $(3,2)$ and $\mathfrak{su}(2)$ $\CN=4$ SYM that will be discussed in much greater detail in \cite{2020}. In particular, we show that the $\CA(6)$ VOA saturates an inequality on the number of strong generators for VOAs in the class of theories discussed in \cite{2020}. We conclude with some open questions and future directions.

\section{The Index relation}\label{indexrel}
In this section we derive the index relation \eqref{indrel} and explain some of its consequences before setting the stage for a discussion of the GVSI in section \ref{GVSI}.

Let us begin by briefly reminding the reader that the Schur index \cite{Gadde:2011uv} is a specialization of the $\CN=2$ superconformal index counting certain, at worst $1/4$-BPS, local operators. In particular, the Schur index is a refined signed trace over the Hilbert space of local operators, $\CH$
\begin{equation}\label{schurdef}
\CI(q;\vec{x}):=\Tr|_{\CH}(-1)^Fe^{-\beta\Delta}q^{E-R}\prod_i(x_i)^{f_i}~,
\end{equation}
where $|q|<1$, $\vec{x}$ is a vector of flavor fugacities corresponding to weights $\vec{f}$, $F$ is fermion number, $E$ is the conformal dimension, $R$ is the $SU(2)_R$ weight, and $\Delta:=\left\{\CQ_{2\dot-},(\CQ^{2\dot-})^{\dagger}\right\}$.

Given this definition, we would like to construct $\CI_{(3,2)}$, $\CI_{\mathfrak{su}(2)}$, and check \eqref{indrel}. For ease of reference in what follows, we reproduce this relation below
\begin{equation}\label{indrel2}
\CI_{(3,2)}(q)=\CI_{\mathfrak{su}(2)}(q^3;q^{1\over2})~,
\end{equation}
where we again remind the reader that \lq\lq$\mathfrak{su}(2)$" stands for $\mathfrak{su}(2)$ $\CN=4$ SYM.

The simple relation in \eqref{indrel2} can, at some level, be anticipated from the simple form the Schur index of $D_3$ takes \footnote{This phenomenon is familiar from the general $D_p(SU(N))$ theories: they are very much the closest AD cousins of free theories \cite{Buican:2017fiq,Buican:2017rya,Buican:2019huq,Buican:2019kba}.}: it has simple \lq\lq single letter" contributions when written as in \cite{Xie:2016evu,kac2017remark,Creutzig:2017qyf}. In particular, we have
\begin{equation}\label{D3ind}
\CI_{D_3}(q;y)={\rm P.E.}\left[\left({q(1+q)\over1-q^3}\right)\chi_{\rm adj}(y)\right]~,
\end{equation}
where $y$ is a fugacity for the $D_3$'s $SU(2)$ flavor symmetry that we gauge to produce the $(3,2)$ SCFT, and we define the \lq\lq plethystic exponential" to be ${\rm P.E.}[g(q;x_1,\cdots,x_p)]:=\exp\left(\sum_{n=1}^{\infty}{1\over n}g(q^n;x_1^n,\cdots,x_p^n)\right)$. Notice that the expression in \eqref{D3ind} is not too different from the index of the hypermultiplets in the SYM theory
\begin{equation}\label{hindex}
\CI_{\rm hyp}(q;y,x)={\rm P.E.}\left[{q^{1\over2}\over1-q}(x+x^{-1})\chi_{\rm adj}(y)\right]~,
\end{equation}
where $y$ is the fugacity for the $\mathfrak{su}(2)$ flavor symmetry we gauge to produce the SYM theory, and $x$ is a fugacity for the remaining $SU(2)_F$ $\CN=2$ flavor symmetry we have discussed at length above.

To generate the relation in \eqref{indrel2} given the building blocks in \eqref{D3ind} and \eqref{hindex}, we need only integrate the vector multiplet contribution to the index 
 \begin{equation}\label{N4}
\CI_{\rm vec}(q;y)={\rm P.E.}\left[-{2q\over1-q}\chi_{\rm adj}(y)\right]~,
\end{equation}
over the $\mathfrak{su}(2)$ Haar measure and include the appropriate matter contributions. Doing so, we find
\begin{equation}\label{32ind}
\CI_{\mathfrak{su}(2)}(q;x)=\int d\mu_{\mathfrak{su}(2)}(y)\cdot\CI_{\rm vec}(q;y)\cdot\CI_{\rm hyp}(q;y,x)~,
\end{equation}
and 
\begin{equation}
\CI_{(3,2)}(q)=\int d\mu_{SU(2)}(y)\cdot\CI_{\rm vec}(q;y)\cdot\left(\CI_{D_3}(q;y)\right)^3~,
\end{equation}
from which \eqref{indrel2} easily follows upon taking $q\to q^3$ and
$x\to q^{1\over2}$ in 
%\eqref{N4}
\eqref{32ind}.

It is interesting to note that one consequence of the index relation we have derived in \eqref{indrel2} is that the $U(1)_r$ anomalies discussed in the introduction must match. Indeed, since the $SU(2)_F$ flavor symmetry has vanishing linear 't Hooft anomaly \footnote{This statement is more generally true of any $\CN=2$ flavor symmetry \cite{Buican:2013ica}.}, we see that $\CI_{(3,2)}$ and $\CI_{\mathfrak{su}(2)}$ have the same \lq\lq high-temperature" behavior and hence the same value of $a-c$ (see the discussion in \cite{Buican:2015ina}). Indeed,
\begin{eqnarray}
a_{(3,2)}&=&c_{(3,2)}=2\ \Rightarrow\ {\rm Tr}|_{(3,2)}\ U(1)_r=0~, \cr a_{\mathfrak{su}(2)}&=&c_{\mathfrak{su}(2))}={3\over4}\ \Rightarrow\ {\rm Tr}|_{\mathfrak{su}(2)}\ U(1)_r=0~.\ \ \ \
\end{eqnarray}

While the above discussion strongly suggests that the Schur sectors of $\mathfrak{su}(2)$ $\CN=4$ SYM and the $(3,2)$ SCFT are related, it is not at all obvious from the facts presented thus far that there is a particularly simple map between these two sectors. Indeed, it turns out that both the $(3,2)$ and $\mathfrak{su}(2)$ $\CN=4$ theories have fermionic and bosonic Schur operators \footnote{This statement can be seen by carefully combining the vector multiplet gaugino Schur operators into composites with the bosonic matter sector Schur operators. This procedure was carried out for the $(3,2)$ theory in \cite{Buican:2016arp}.}. Therefore, even though the Schur indices are closely related, the $(-1)^F$ in the definition \eqref{schurdef} can sweep various differences in operator content under the rug. For example, at the level of the index, there is no general way to distinguish operator relations amongst bosons from fermionic Schur operators with the same quantum numbers (and vice versa). The main result of the next section will be to show that, in spite of this possibility, there is in fact a GVSI between the two Schur sectors when analyzed at the levels of the corresponding VOAs.

\section{The exact graded vector space isomorphism}\label{GVSI}
To gain further insight into the mechanism that explains the index relation in \eqref{indrel2}, it is useful to consider the associated 2D VOAs in the sense of \cite{Beem:2013sza}. Indeed, for any 4D $\CN=2$ SCFT, $\CT$, \cite{Beem:2013sza} showed there is a corresponding 2D chiral algebra, $\chi[\CT]$, living on a plane $\mathcal{P}\simeq\mathbb{R}^2\subset\mathbb{R}^4$. Via an $SU(2)_R$ twisting on $\mathcal{P}$, one can show that each Schur operator resides in a cohomology class (with respect to a nilpotent supercharge) that is in one-to-one correspondence with a state of the VOA living on $\mathcal{P}$.

We can make contact with the discussion of the previous section by noting that the torus partition function for $\chi[\CT]$ takes the form
\begin{equation}\label{ZT2def}
Z_{T^2;\chi[\CT]}(q;z;\vec{x})=\Tr\ z^{M^{\perp}}q^{L_0-{c_{2d}\over24}}\prod_i(x_i)^{f_i}~,
\end{equation}
where $M^{\perp}=j_1-j_2=-r$ is the spin transverse to $\mathcal{P}$, $c_{2d}=-12c_{4d}$ is central charge of the VOA in terms of the 4D central charge, $L_0$ is the holomorphic weight, and $x_i$ is a flavor fugacity (sometimes referred to in the VOA literature as a Jacobi variable) with corresponding weight $f_i$. Note here that $r$ is just the $U(1)_r$ charge and that it is a conserved (though non-local) charge of the VOA.

Since they count essentially the same states, it should come as no surprise that \cite{Beem:2013sza}
\begin{equation}\label{2dindex}
Z_{T^2;\chi[\CT]}(q;(-1);\vec{x})=q^{-{c_{2d}\over24}}\CI_{\CT}(q;\vec{x})~.
\end{equation}
In this relation, the holomorhpic dimension maps as follows: $h=E-R$. Therefore, we can translate \eqref{indrel2} into the 2D statement that
\begin{equation}\label{Zrel}
Z_{T^2,\chi[(3,2]}(q;(-1))=q^{1\over8}Z_{T^2,\chi[\mathfrak{su}(2)]}(q^3;(-1);q^{1\over2})~.
\end{equation}

However, our goal is to go beyond this relation and to understand if there is a non-trivial mapping of states in the two VOAs. On the $\CN=4$ side we have the following:
\begin{enumerate}
\item{From \cite{Beem:2013sza,Bonetti:2018fqz}, we know that
     $\chi[\mathfrak{su}(2)]={\rm sVir}_{{\rm sm}\ \CN=4}^{c=-9}$, i.e.,
     the VOA associated with the $\mathfrak{su}(2)$ $\CN=4$ theory is
     the small $\CN=4$ super-Virasoro algebra
 at $c=-9$.}
\item{The bosonic strong generators of this VOA are the three affine currents of $\widehat{\mathfrak{su}(2)}_{-{3\over2}}$
\begin{equation}
{\rm Bos.\ strong\ gens.:}\ J^{0,\pm}\in\widehat{\mathfrak{su}(2)}_{-{3\over2}}\subset{\rm sVir}_{{\rm sm}\ \CN=4}^{c=-9}~.
\end{equation}
These currents have $h=1$, $r=0$, and are related to 4D Higgs branch operators (i.e., moment map primaries of $\widehat{\CB}_1$ type with $R=1$ in the language of \cite{Dolan:2002zh} that correspond to $SU(2)_F$; see also \cite{Dobrev:1985qv}). Therefore $J^{0,\pm}$ are also the bosonic generators of the HL chiral ring. Here, $T$ is not an independent generator (it is the Sugawara stress tensor).}
\item{The fermionic strong generators are the four $\CN=4$ supercurrents of $h={3\over2}$
\begin{equation}
{\rm Ferm.\ strong\ gens.:}\ G^{\pm}, \ \tilde G^{\pm}~,
\end{equation}
where $G^{\pm}$ have $r=1/2$ and $\tilde G^{\pm}$ have $r=-1/2$. These latter currents are also HL generators (they reside in multiplets of type $\bar\CD_{{1\over2}(0,0)}$) and the former are not (they reside in multiplets of type $\CD_{{1\over2}(0,0)}$).
}
\end{enumerate}

On the $(3,2)$ side we have the following:
\begin{enumerate}
\item{From \cite{Buican:2016arp}, we know that $\chi[(3,2)]=\CA(6)$, i.e., the VOA associated with the $(3,2)$ theory is the $c=-24$ $\CA(6)$ chiral algebra of Feigin, Feigin, and Tipunin \cite{Feigin:2007sp,Feigin:2008sg}.}
\item{The only bosonic strong generator of this VOA is the energy momentum tensor
\begin{equation}\label{bgen}
{\rm Bos.\ strong\ gens.:}\ T\in{\rm Vir}_{c=-24}\subset \CA(6)~.
\end{equation}
This operator has $h=2$, $r=0$, and is related to the 4D $SU(2)_R$ current (it is a superconformal descendant in the $\widehat{\CC}_{0(0,0)}$ stress tensor multiplet). This is not a Higgs branch or HL operator.}
\item{The fermionic currents are two $h=4$ currents
\begin{equation}\label{Fgen}
{\rm Ferm.\ strong\ gens.:}\ \Psi, \ \tilde \Psi~,
\end{equation}
where $\Psi$ has $r=1/2$ and $\tilde\Psi$ has $r=-1/2$. Using the $D_3$ Macdonald index \cite{Buican:2015tda} or conformal perturbation theory, one can argue that these operators have to be of $\widehat{\CC}_{{3\over2}({1\over2},0)}\oplus\widehat{\CC}_{{3\over2}(0,{1\over2})}$ type \cite{Buican:2016arp}. Hence, as discussed in the introduction, the $(3,2)$ theory has no Higgs branch or HL operators \footnote{Consistency with the bosonic generator in \eqref{bgen} suggests that these operators cannot come from 4D superconformal descendants of $\CD_{R(0,j_2)}\oplus\overline{\CD}_{R(j_1,0)}$ multiplets. Indeed, otherwise we would find the HL ring only contains fermionic generators whereas all examples of HL rings we are aware of (for interacting theories) contain bosonic generators as well. It would be interesting to see if one can prove a theorem forbidding HL rings with purely fermionic generators for interacting theories.}.
}
\end{enumerate}

Clearly the above VOAs are not isomorphic: the central charges and number of strong generators are different. Moreover the HL rings and Higgs branches do not match. Still, given the result in \eqref{Zrel} we can hope for a non-trivial isomorphism of the VOAs when thought of as $(-1)^F$ (or equivalently $U(1)_r$) graded vector spaces. In particular, we would like to see if we can map operators to operators and null states to null states while preserving statistics and $U(1)_r$. In other words, we would like to see if we can construct the GVSI promised in the introduction. Such a GVSI implies \eqref{indrel2} and \eqref{Zrel}, but, as discussed in section \ref{indexrel}, it is a much stronger result.

To motivate such a GVSI, let us work out what such a map
\begin{equation}\label{varphimap}
\varphi: \chi[\mathfrak{su}(2)]\to\chi[(3,2)]~,
\end{equation}
would look like for small values of $h$. Crucially, in addition to preserving $U(1)_r$, we are forced by \eqref{Zrel} to set
\begin{equation}\label{hrel}
h_{(3,2)}=3h_{\mathfrak{su}(2)}+{1\over2}f~,
\end{equation}
where $h_{(3,2)}$ is the holomorphic dimension in $\CA(6)$, $h_{\mathfrak{su}(2)}$ is the corresponding quantity in ${\rm sVir}_{{\rm sm}\ \CN=4}^{c=-9}$, and $f$ is the weight under $SU(2)_F$.

The lowest-dimensional non-trivial operator in $\CA(6)$ is the stress tensor at $h_{(3,2)}=2$. The constraint \eqref{hrel} fixes
\begin{equation}\label{Jm}
\varphi(J^-)=T~.
\end{equation}
The next non-trivial state is $\partial T$ at $h_{(3,2)}=3$ , and again \eqref{hrel} fixes a unique choice
\begin{equation}\label{J0}
\varphi(J^0)=\partial T~. 
\end{equation}
At $h_{(3,2)}=4$ we have two bosonic states: $\partial^2T$ and $T^2$. At the level of a GVSI, we can set $\varphi(J^+)$ and $\varphi((J^-)^2)$ to any two independent linear combinations of these states. However, it is natural to also demand that the normal-ordered product is respected so that
\begin{equation}\label{Jp}
\varphi(J^+)=\partial^2T~, \ \varphi((J^-)^2)=T^2~.
\end{equation}
We also have two $h_{(3,2)}=4$ fermionic operators: $\Psi$ and $\tilde\Psi$. These are uniquely identified via the requirement that $\varphi$ respect $U(1)_r$ as
\begin{equation}\label{Gpm}
\varphi(G^-)=\Psi~, \ \varphi(\tilde G^-)=\tilde\Psi~.
\end{equation}

The relations in \eqref{Jm}, \eqref{J0}, \eqref{Jp}, and \eqref{Gpm} succinctly express the idea behind distorted symmetry breaking: the VOA generators that map to HL states on the SYM Higgs branch (i.e., the $\CN=2$ NG multiplet for the $SU(2)_F$ symmetry breaking and its $\CN=4$ vector multiplet partner) are set to zero in the (trivial) $(3,2)$ HL ring (as in \eqref{distorted}). However, unlike the case of motion onto the Higgs branch, the HL operators are mapped to non-trivial but non-HL states in the $(3,2)$ theory.

Before moving on to a proof, it is also worth considering operators with $h_{(3,2)}=5$ in order to understand how derivatives get mapped by $\varphi$ and to understand the mapping of the remaining fermionic generators of ${\rm sVir}_{{\rm sm}\ \CN=4}^{c=-9}$. To that end, the bosonic operators at $h_{(3,2)}=5$ are $\partial^3 T$ and $T\partial T$. We can again choose $\varphi$ to preserve the normal ordered product by taking
\begin{equation}\label{h5b}
\varphi(\partial J^-)=\partial^3J^-~, \ \varphi(J^0J^-)=(\partial T)T~.
\end{equation}
The fermionic operators at $h_{(3,2)}=5$ are $G^+$ and $\tilde G^+$. The fact that $\varphi$ respects $U(1)_r$ means that
\begin{equation}
\varphi(G^+)=\partial\Psi~, \ \varphi(\tilde G^+)=\partial\tilde\Psi~.
\end{equation}

Finally, it is worth further motivating our proof by considering the fermionic operators with $h_{(3,2)}=6$, since this is the first level with null vectors in $\CA(6)$ \cite{Feigin:2007sp,Feigin:2008sg}
\begin{equation}\label{nullF}
\kappa\partial^2\Psi+T\Psi=\kappa\partial^2\tilde\Psi+T\tilde\Psi=0~, \ \kappa\ne0~.
\end{equation}
These two null vectors are important for our candidate GVSI, $\varphi$, to work. Indeed, \eqref{nullF} means that there are now just two fermionic states at $h_{(3,2)}=6$: $T\Psi\sim\partial^2\Psi$ and $T\tilde\Psi\sim\partial^2\tilde\Psi$. This number is just right because there are only two candidate states in ${\rm sVir}_{{\rm sm}\ \CN=4}^{c=-9}$ that can map onto these $\CA(6)$ fermions, and $U(1)_r$ fixes this mapping uniquely
\begin{equation}
\varphi(J^-G^-)=T\Psi~, \ \varphi(J^-\tilde G^-)=T\tilde\Psi~.
\end{equation}
In addition to these tests, we also explicitly checked the remaining states at $h_{(3,2)}=6$ as well as the states with $h_{(3,2)}=7,8$ and found that they are all consistent with the existence of the above GVSI \footnote{We thank J.~Shafiq for rechecking some of these results.}.

This discussion makes plausible the following theorem:

\noindent
{\bf Theorem 1:} The map $\varphi$ in \eqref{varphimap} is a GVSI respecting \eqref{hrel} with the following additional properties:
\begin{enumerate}
 \item{$\varphi$ respects $U(1)_r$ charge: $r(\CO)=r(\varphi(\CO))$ for all $\CO\in\chi[\mathfrak{su}(2)]$. By the taxonomy of Schur operators, this means $\varphi$ respects the Bose/Fermi statistics of the operators it maps.}
 \item{There exists a basis $\{\mathcal{O}_i\}$ of $\chi[\mathfrak{su}(2)]$ such that every $\mathcal{O}_i$ is a normal ordered product of strong generators and/or their derivatives, and $\varphi(\mathcal{O}_i) =\varphi(\mathcal{O}_{i,1})\cdots\varphi(\mathcal{O}_{i,k_i})$ when $\mathcal{O}_i = \mathcal{O}_{i,1}\cdots \mathcal{O}_{i,k_i}$ \footnote{By definition of strong generators, every element of $\chi[\mathfrak{su}(2)]$ is written as a linear combination of normal ordered products of strong generators and/or their derivatives. However, non-trivial null operator relations imply that some of these normal ordered products are linearly dependent. Our statement here is that there exists a set $\{\mathcal{O}_i\}$ of linearly independent normal ordered products for which $\varphi$ preserves the structure of the normal ordering.}.}
  \item{$\varphi(\partial^k \mathcal{O})=\partial^{3k}\varphi(\mathcal{O})$ for all strong generators $\mathcal{O}$ of $\chi[\mathfrak{su}(2)]$ in the basis mentioned in property 2.}
\end{enumerate}

The key idea that leads to a proof of the theorem is to compare a decomposition of ${\rm sVir}_{{\rm sm}\ \CN=4}^{c=-9}$ in terms of Weyl modules of the universal affine vertex algebra, $V_{-{3\over2}}(\mathfrak{su}(2))$, given in \cite{Creutzig:2018ltv} with a corresponding decomposition of the $\CA(6)$ VOA in terms of Virasoro modules given in \cite{Feigin:2007sp}.

In light of this discussion, the following lemma will be useful in proving theorem 1:

\noindent
{\bf Lemma 2:} Under the identification in \eqref{hrel}, we have that 
\begin{equation}
M_{m+1,1;6}\simeq V_{-{3\over2}}(m\omega)~,
\end{equation}
as a linear equivalence of graded vector spaces. Here $M_{m+1,1;6}$ is the Virasoro module with Kac labels $(m+1,1)$ at $c=-24$, and $V_{-{3\over2}}(m\omega)$ is the Weyl module of $V_{-{3\over2}}(\mathfrak{su}(2))$ associated with $m\omega$ (where $\omega$ is the fundamental weight of $\mathfrak{su}(2)$).

\noindent
{\bf Proof (Lemma 2):} As discussed in \cite{Creutzig:2018ltv}, $V_{-{3\over2}}(m\omega)$ has no singular vectors and is therefore spanned by all vectors of the form
\begin{eqnarray}\label{Jspace}
&&|s,\left\{a_k\right\},\left\{b_k\right\},\left\{c_k\right\}\rangle_{\mathfrak{su}(2)}:=\cr&&\left(\prod_{k=1}^{\infty}(J^+_{-k})^{a_k}(J^0_{-k})^{b_k}(J^-_{-k})^{c_k}\right)|s\rangle_{\mathfrak{su}(2)}~,
\end{eqnarray}
where the product is taken so that $J^A_{-k}$ is on the right of
$J^B_{-\ell}$ if $k<\ell$, $a_k$, $b_k$, and $c_k$ are non-negative
integers, and $s=0,\cdots,m$ labels eigenstates of $J^0_0$ with
eigenvalue $s-\frac{m}{2}$ \footnote{Note that $f$ in \eqref{hrel} is twice
the eigenvalue of $J^0_0$.}. Therefore, $V_{-{3\over2}}(m\omega)$ is {\it linearly} isomorphic to the Virasoro Verma module spanned by
\begin{eqnarray}\label{Lspace}
&&|s,\left\{a_k\right\},\left\{b_k\right\},\left\{c_k\right\}\rangle_{\rm Vir}:=\cr&&\left(\prod_{k=1}^{\infty}(L_{-3k-1})^{a_k}(L_{-3k})^{b_k}(L_{-3k+1})^{c_k}\right)L_{-1}^s|h\rangle~,\ \ \ \
\end{eqnarray}
where we set
\begin{eqnarray}\label{hsu2}
h&=&3{(m\omega,m\omega+2\rho)\over2(k_{\mathfrak{su}(2)}+h^{\vee}_{\mathfrak{su}(2)})}-{m\over2}\cr&=&{m(3m+5)\over2}~,
\end{eqnarray}
in order to guarantee that $|s\rangle_{\mathfrak{su}(2)}\to(L_{-1}^s)|h\rangle$ under our identification \eqref{hrel}. We also have the following mapping of modes that is manifestly compatible with \eqref{hrel}
\begin{equation}\label{JLmap}
J_{-k}^{\pm}\to L_{-(3k\pm1)}~, \ J_{-k}^0\to L_{-3k}~.
\end{equation}
We can write the space spanned by our states in \eqref{Lspace} more succinctly as the quotient of the Verma module, ${\rm Verma}(h)$, by the subspace generated by $(L_{-1})^{m+1}|h\rangle$, and so we find
\begin{equation}\label{affine}
V_{-{3\over2}}(m\omega)\simeq{\rm Verma}(h)/{\rm
 Vir}_-(L_{-1})^{m+1}|h\rangle~,
\end{equation}
where ${\rm Vir}_-$ is the subalgebra spanned by $L_{-k}$ with $k>0$.

Let us now analyze $M_{m+1,1;6}$. As discussed in \cite{Feigin:2007sp}, this is the quotient of ${\rm Verma}(h_{m+1,1})$ by a subspace generated by a singular vector $|\psi\rangle$ at level $m+1$
\begin{equation}\label{minimal}
M_{m+1,1;6}\simeq{\rm Verma}(h_{m+1,1})/{\rm Vir}_-|\psi\rangle~,
\end{equation}
where
\begin{equation}
h_{m+1,1}={(6(m+1)-1)^2-25\over24}={m(3m+5)\over2}~.
\end{equation}
Note that this holomorphic dimension coincides with the one in \eqref{hsu2}. Therefore, we have our result. $\square$

With this lemma in hand, we now turn to the proof of the main theorem:

\noindent
{\bf Proof (Theorem 1):} We have from \cite{Creutzig:2018ltv}
\begin{equation}\label{N4VOA}
{\rm sVir}_{{\rm sm}\
 \CN=4}^{c=-9}\simeq\bigoplus_{m=0}^{\infty}\left(\pi_{m+1}\otimes
					     V_{-{3\over2}}(m\omega)\right)~,
\end{equation}
where $\pi_k$ is the $k$-dimensional irreducible representation of
$\mathfrak{su}(2)$ and the rest of the quantities are defined as in
lemma 2. 

Similarly, we have from \cite{Feigin:2007sp}
\begin{equation}\label{A6VOA}
\CA(6)\simeq\bigoplus_{m=0}^{\infty}\left(\pi_{m+1}\otimes M_{m+1,1;6}\right)~,
\end{equation}
where $\pi_{m+1}$ is defined as in \eqref{N4VOA}, and the rest of the
quantities are defined as in lemma 2. Finally, from lemma 2 we have that
$V_{-{3\over2}}(m\omega)\simeq M_{m+1,1;6}$ as graded linear spaces. 

In the rest of this proof, we show that the above graded linear isomorphism can be
equipped with
the additional properties listed in Theorem 1. 
As shown in the proof of Theorem 2.5 of
\cite{Creutzig:2018ltv}, the highest weight state of $\pi_{m+1}\otimes V_{-\frac{3}{2}}(m\omega)$ in
\eqref{N4VOA} corresponds to $G^+\partial G^+\partial^2G^+\cdots
\partial^{m-1}G^+$ \footnote{Here the ``highest weight'' is in the sense
of $\mathfrak{su}(2)\otimes V_{-\frac{3}{2}}(\mathfrak{su}(2))$.}, where the
$\mathfrak{su}(2)$ action is such that
$(G^+,\tilde{G}^+)$ transforms as a
doublet.
Therefore the decomposition \eqref{N4VOA}
implies that ${\rm sVir}_{{\rm sm}\ \CN=4}^{c=-9}$ is spanned by the
normal ordered products
\begin{align}\label{span1}
\prod_{k=1}^\infty(\partial^{k-1}J^+)^{a_k}&(\partial^{k-1}J^0)^{b_k}(\partial^{k-1}J^-)^{c_k}
\nonumber\\[-3mm]
&G^{(s_1}_{(i_1}\partial G^{s_2}_{i_2} \cdots \partial^{m-1}G^{s_m)}_{i_m)}~,
\end{align}
where $G^0_1
:= G^-,\,G^{1}_1:= G^+,\,G^0_2 := \tilde{G}^-,\,G^1_2 :=
\tilde{G}^+,\, s_k\in\{0,1\}, i_k\in\{1,2\}$, and we used the state operator map $J^{A}_{-k}|0\rangle \to
\partial^{k-1}J^{A}(0)$ for $A=0,\pm$ and $k\geq 1$. Note that both the
sub-scripts and super-scripts of $G^s_i$ are completely
symmetric in \eqref{span1}.

We now turn to the $\mathcal{A}(6)$ side. We use the free field
realization of $\mathcal{A}(6)$ discussed in \cite{Feigin:2007sp}; $T=\frac{1}{2}(\partial\phi)^2 +
 \frac{5}{2\sqrt{3}}\partial^2\phi,\,\Psi=e^{-\sqrt{3}\phi}$ and
 $\tilde{\Psi}=
[\frac{1}{2\pi i}\oint dz\,e^{\sqrt{3}\phi(z)},e^{-\sqrt{3}\phi}]$ where $\phi$
 is a free field such that $\phi(z)\phi(0)\sim \log z$. With this
 realization, the highest weight state of $\pi_{m+1}\otimes M_{m+1,1;6}$ in \eqref{A6VOA}
 is written as $e^{-m\sqrt{3}\phi}
 \propto \Psi\partial^3\Psi \partial^6\Psi\cdots
 \partial^{3m-3}\Psi$ \footnote{
%Here $e^{-m\sqrt{3}\phi}$ is
%a particular case of the primary vertex operator
% $V_{r,s}(z)=e^{\left(-(r-1)\sqrt{3}+(s-1)\frac{1}{2\sqrt{3}}\right)\phi(z)}$ in the $(1,6)$
% minimal model. 
One can show that $e^{-m\sqrt{3}\phi}\propto\Psi\partial^3\Psi \partial^6\Psi\cdots
 \partial^{3m-3}\Psi$ as follows. First note that 
$e^{-\sqrt{3}\phi(z_1)}\cdots e^{-\sqrt{3}\phi(z_m)} =
 e^{-\sqrt{3}(\phi(z_1)+\cdots +\phi(z_m))} \prod_{1\leq
 i<j\leq m}(z_i-z_j)^{3}$. Differentiating this
 identity and using
 $\Psi=e^{-\sqrt{3}\phi}$, we see that
 $\Psi(z_1)\partial^3\Psi(z_2)\partial^6\Psi(z_3)\cdots
 \partial^{3m-3}\Psi(z_m) \propto e^{-m\sqrt{3}\phi(z_m)} + X$, where $X$
 vanishes in the limit of $z_k \to z_m$ for $k=1,\cdots,m-1$. This
 implies that the normal ordered product $\Psi\partial^3\Psi\cdots
 \partial^{3m-3}\Psi$ is proportional to $e^{-m\sqrt{3}\phi}$.}, where the
$\mathfrak{su}(2)$-action is such that $(\Psi,\tilde{\Psi})$ transforms
as a doublet. Then \eqref{A6VOA} and 
%the isomorphism between \eqref{affine} and \eqref{minimal} 
Lemma 2
imply that
$\mathcal{A}(6)$ is spanned by
\begin{align}\label{span2}
 \prod_{k=1}^\infty (\partial^{3k-1} T)^{a_k} &(\partial^{3k-2}T)^{b_k}
 (\partial^{3k-3}T)^{c_k}
\nonumber\\[-3mm]
&\partial^{s}\Big(\Psi_{(i_1}\partial^3\Psi_{i_2}\cdots \partial^{3m-3}\Psi_{i_m)}\Big)
\end{align}
where $\Psi_1\equiv \Psi,\,\Psi_2\equiv \tilde{\Psi},\, s\in\{0,\cdots,m\},\,i_k\in
\{1,2\}$, and we used the state operator map $L_{-k}|0\rangle \to
\partial^{k-2}T(0)$ for $k\geq 2$.

Let us now consider the
linear map $\varphi: {\rm sVir}_{{\rm sm}\ \CN=4}^{c=-9}
\to \mathcal{A}(6)$ that satisfies the properties
listed in Theorem 1 for the basis \eqref{span1} of ${\rm sVir}_{\rm
sm\ \CN=4}^{c=-9}$.
Here we set $\varphi(J^-) = T,\,\varphi(J^0) = \partial T,\,\varphi(J^+)
= \partial^2 T,\,\varphi(G^-) = \Psi,\,\varphi(G^+) =
\partial\Psi,\,\varphi(\tilde{G}^-) =
\tilde{\Psi}$ and $\varphi(\tilde{G}^+) = \partial \tilde{\Psi}$. 
%We
% will show below that this $\varphi$ is a graded linear isomorphism.
% Note that
Then  $\varphi$ maps the basis \eqref{span1} to
\begin{align}\label{span3}
  \prod_{k=1}^\infty &(\partial^{3k-1} T)^{a_k} (\partial^{3k-2}T)^{b_k}
 (\partial^{3k-3}T)^{c_k}
\nonumber\\[-2mm]
&\frac{1}{m!}\sum_{\sigma \in S_m}\partial^{s_{\sigma(1)}}\Psi_{(i_1}\,\partial^{3+s_{\sigma(2)}}\Psi_{i_2}\,\cdots \partial^{3m-3+s_{\sigma(m)}}\Psi_{i_m)}~,
\end{align}
where $S_m$ is the symmetric group of degree $m$ (note from comparing \eqref{span1} and \eqref{span3}, properties 1-3 mentioned in the theorem follow). Below we show that this
$\varphi$ is a graded linear isomorphism. It is straightforward to show that $\varphi$ is
compatible with \eqref{hrel} and therefore is a graded linear
map. Then all we need to show is that \eqref{span2} and
\eqref{span3} span the same space so that $\varphi$ is a linear
isomorphism. 

The fact that \eqref{span2} and \eqref{span3} span the same space can be seen
as follows. We focus on the sub-space spanned by the $\mathfrak{su}(2)$ highest
weight states, since its orthogonal complement is generated by the
action of
the $\mathfrak{su}(2)$ lowering generator. We start with the fact that 
$\partial^{s}(\Psi \partial^3\Psi \cdots \partial^{3m-3}\Psi) \propto
\partial^s e^{-m\sqrt{3}\phi} = P_s(\partial \phi)e^{-m\sqrt{3}\phi}$,
where $P_s(\partial \phi)$ is a degree-$s$ differential polynomial of
$\partial\phi$ involving $s$ derivatives. Using $T=\frac{1}{2}(\partial \phi)^2 +
\frac{5}{2\sqrt{3}}\partial^2\phi$, we see that
$P_s(\partial \phi)e^{-m\sqrt{3}\phi} \propto
(\partial\phi)^se^{-m\sqrt{3}\phi} +
\sum_{k=1}^{s}Q_k(T)(\partial\phi)^{s-k}e^{-m\sqrt{3}\phi}$, where $Q_k(T)$ are differential
polynomials of $T$. This implies that \eqref{span2} for $a_k,b_k,c_k\geq
0$ and $0\leq s\leq m$ span the same space
as 
\begin{align}\label{span4}
\prod_{k=1}^\infty (\partial^{3k-1} T)^{a_k} (\partial^{3k-2}T)^{b_k}
 (\partial^{3k-3}T)^{c_k}\;\,(\partial\phi)^s e^{-m\sqrt{3}\phi}~,
\end{align}
for the same ranges of $a_k,b_k,c_k$ and $s$.
Similarly, one can show that \eqref{span3} for $a_k,b_k,c_k\geq 0$ and
$s_k\in \{0,1\}$ span the same space as
\eqref{span4} where we identify $s = s_1 + \cdots + s_m$
\footnote{Indeed, since $e^{-\sqrt{3}\phi(z_1)}\cdots
e^{-\sqrt{3}\phi(z_m)} = e^{-\sqrt{3}(\phi(z_1)+\cdots+\phi(z_m))}\prod_{1\leq i<j\leq
m}(z_i-z_j)^3$, it is straight forward to show
that $\partial^{s_1}\Psi\partial^{3+s_2}\Psi \cdots \partial^{3m-3+s_m}\Psi$
is written as $P_{s_1+\cdots+s_m}(\partial
\phi)e^{-m\sqrt{3}\phi}$ for a differential polynomial
$P_{s_1+\cdots+s_m}(\partial \phi)$. Since $s_k\in\{0,1\}$, $s_1+\cdots+s_m \in
\{0,\cdots,m\}$. Therefore, the same argument as above implies
\eqref{span3} for $a_k,b_k,c_k\geq 0$ and $s_k \in \{0,1\}$ span the
same space as \eqref{span4}.}. Hence, the spaces spanned by \eqref{span2} and \eqref{span3}
are identical, and therefore $\varphi$ is
a graded linear isomorphism. $\square$

\section{$|\chi[(3,2)]|$ versus $|\chi[\mathfrak{su}(2)]|$ and an inequality}\label{ineq}
In this section we will describe some additional relations between the $(3,2)$ VOA and its $\CN=4$ SYM VOA cousin. Our starting point is to recall that the number of strong generators in $\CA(6)=\chi[(3,2)]$ saturates a universal bound on the number of strong generators in VOAs related to 4D $\CN=2$ theories with exactly marginal gauge couplings \cite{Buican:2016arp}
\begin{equation}
|\chi[\CT]|\ge3~,
\end{equation}
where $\CT$ is any 4D $\CN=2$ SCFT with an exactly marginal gauge coupling, and $|\chi[\CT]|$ denotes the number of strong generators of $\chi(\CT)$ (here $|\CA(6)|=3$ since, as discussed in the previous section, it is strongly generated by $T$, $\Psi$, and $\tilde\Psi$).

In what follows, we would like to argue that $|\CA(6)|$ saturates another bound on the number of strong generators, but this time from below.

However, to discuss this bound, we will need to tease some results that will appear in our upcoming work \cite{2020}. In particular, in \cite{2020} we will argue that one can generalize some of the above results (and find various new ones) by considering an infinite set of $\CN=2$ SCFTs that are closely related to the $(3,2)$ theory.

\begin{figure}
 \begin{center}
\begin{tikzpicture}[place/.style={circle,draw=blue!50,fill=blue!20,thick,inner sep=0pt,minimum size=6mm},transition2/.style={rectangle,draw=black!50,fill=red!20,thick,inner sep=0pt,minimum size=8mm},auto]

 \node[place] (1) at (-4,0) [shape=circle] {\;$N$\;};
 \node[transition2] (2) at (-4-1,0) {\;$D_{2,N}$\;} edge (1);
 \node[transition2] (3) at (-4+1,0) {\;$D_{2,N}$\;} edge (1);
 \node[transition2] (4) at (-4,1) {\;$D_{2,N}$\;} edge (1); 
 \node[transition2] (5) at (-4,-1) {\;$D_{2,N}$\;} edge (1);
 \node at (-4,-2) {$(2,N)$}; 

\end{tikzpicture}
\caption{We may further generalize the $(3,2)$ theory by considering a diagonal exactly marginal $SU(N)$ gauging of four $D_2(SU(N)):=D_{2,N}$ theories. Here ${\rm gcd}(2,N)=1$.}
\label{fig:2N}
 \end{center}
\end{figure}
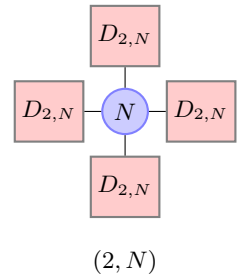

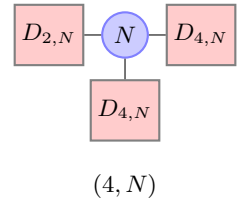
\begin{figure}
 \begin{center}
\begin{tikzpicture}[place/.style={circle,draw=blue!50,fill=blue!20,thick,inner sep=0pt,minimum size=6mm},transition2/.style={rectangle,draw=black!50,fill=red!20,thick,inner sep=0pt,minimum size=8mm},auto] 

 \node[place] (1) at (4,0) [shape=circle] {\;$N$\;};
 \node[transition2] (2) at (4-1,0) {\;$D_{2,N}$\;} edge (1);
 \node[transition2] (3) at (4+1,0) {\;$D_{4,N}$\;} edge (1);
 \node[transition2] (4) at (4,-1) {\;$D_{4,N}$\;} edge (1); 

 \node at (4,-2) {$(4,N)$}; 

  \end{tikzpicture}
\caption{We may also generalize the $(3,2)$ theory by considering a diagonal exactly marginal $SU(N)$ gauging of two $D_4(SU(N)):=D_{4,N}$ theories and one $D_2(SU(N)):=D_{2,N}$ SCFT. Here ${\rm gcd}(4,N)=1$.}
\label{fig:4N}
 \end{center}
\end{figure}

\begin{figure}
 \begin{center}
\begin{tikzpicture}[place/.style={circle,draw=blue!50,fill=blue!20,thick,inner sep=0pt,minimum size=6mm},transition2/.style={rectangle,draw=black!50,fill=red!20,thick,inner sep=0pt,minimum size=8mm},auto]

 \node[place] (1) at (8,0) [shape=circle] {\;$N$\;};
 \node[transition2] (2) at (8-1,0) {\;$D_{2,N}$\;} edge (1);
 \node[transition2] (3) at (8+1,0) {\;$D_{3,N}$\;} edge (1);
 \node[transition2] (4) at (8,-1) {\;$D_{6,N}$\;} edge (1); 

 \node at (8,-2) {$(6,N)$}; 

  \end{tikzpicture}
\caption{As a final generalization of the $(3,2)$ theory we may considering a diagonal exactly marginal $SU(N)$ gauging of a $D_2(SU(N)):=D_{2,N}$ theory, a $D_3(SU(N)):=D_{3,N}$ SCFT, and a $D_6(SU(N)):=D_{6,N}$ theory. Here ${\rm gcd}(6,N)=1$.}
\label{fig:6N}
 \end{center}
\end{figure}
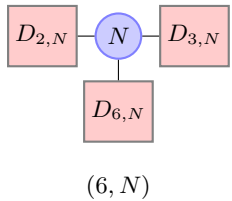

We call these generalized theories $(n,N)$ SCFTs (they have also been studied in \cite{DelZotto:2015rca} under different names). We consider four infinite cases:
\begin{enumerate}
\item{The $(3,N)$ SCFT with ${\rm gcd}(3,N)=1$ as in figure \ref{fig:3N}. It consists of an exactly marginal $SU(N)$ gauging of three $D_3(SU(N)):=D_{3,N}$ SCFTs. The $(3,2)$ SCFT discussed in the previous sections has $N=2$ (and $D_{3,2}:=D_3$).}
\item{The $(2,N)$ SCFT with ${\rm gcd}(2,N)=1$ as in figure \ref{fig:2N}. It consists of an exactly marginal $SU(N)$ gauging of four $D_2(SU(N)):=D_{2,N}$ SCFTs.}
\item{The $(4,N)$ SCFT with ${\rm gcd}(4,N)=1$ as in figure \ref{fig:4N}. It consists of an exactly marginal $SU(N)$ gauging of two $D_4(SU(N)):=D_{4,N}$ SCFTs and one $D_{2,N}$ theory.}
\item{The $(6,N)$ SCFT with ${\rm gcd}(6,N)=1$ as in figure \ref{fig:6N}. It consists of an exactly marginal $SU(N)$ gauging of one $D_{2,N}$ SCFT, one $D_{3,N}$ theory, and one $D_6(SU(N)):=D_{6,N}$ SCFT.}
\end{enumerate}
Among other results, we will argue in \cite{2020} for the following generalization of \eqref{indrel}, where $\mathfrak{su}(2)$ $\CN=4$ SYM is generalized to $\mathfrak{su}(N)$ $\CN=4$ SYM
\begin{equation}
\CI_{(n,N)}(q)=\CI_{\mathfrak{su}(N))}(q^n,q^{n/2-1})~,
\end{equation}
and the following generalization of \eqref{hrel}
\begin{equation}\label{hrelnN}
h_{(n,N)}=n\cdot h_{\mathfrak{su}(N)}+\left({n\over2}-1\right)f~.
\end{equation}
Generalizing section \ref{GVSI}, we will argue in \cite{2020} that, among other things, there is an underlying GVSI
\begin{equation}
\varphi_{n,N}:\chi[\mathfrak{su}(N)]\to\chi[(n,N)]~,
\end{equation}
that respects $U(1)_r$ and \eqref{hrelnN}.

\begin{figure}
 \begin{center}
\begin{tikzpicture}[place/.style={circle,draw=blue!50,fill=blue!20,thick,inner sep=0pt,minimum size=6mm},transition2/.style={rectangle,draw=black!50,fill=red!20,thick,inner sep=0pt,minimum size=8mm},auto]

 \node[place] (1) at (0,0) [shape=circle] {\;$N$\;};
 \node[transition2] (2) at (-1,0) {\;$D_{3,N}$\;} edge (1);
 \node[transition2] (3) at (1,0) {\;$D_{3,N}$\;} edge (1);
 \node[transition2] (4) at (0,-1) {\;$D_{3,N}$\;} edge (1); 
 \node at (0,-2) {$(3,N)$}; 
 
 \end{tikzpicture}
\caption{Our main theory of interest, $(3,2)$, has $N=2$ and is the simplest member of this larger class of theories (see figure \ref{fig:32}). More generally, we may consider exactly marginal diagonal $SU(N)$ gaugings of three $D_3(SU(N)):=D_{3,N}$ SCFTs (where $D_{3,2}:=D_3$). The only constraint on $N$ is that ${\rm gcd}(3,N)=1$.}
\label{fig:3N}
 \end{center}
\end{figure}

Given this picture, we claim that $\CA(6)$ saturates a new bound (this time from below) on the number of strong VOA generators in the infinite set of $(n,N)$ SCFTs relative to the number of strong generators in the corresponding $\CN=4$ theory:

\noindent
{\bf Claim 3:} Assuming the conjecture in \cite{Beem:2013sza} for the VOA corresponding to $\mathfrak{su}(N)$ $\CN=4$ SYM, $\chi[\mathfrak{su}(N)]$, \footnote{Actually, we will only need to assume that the low-lying states of the conjectures in \cite{Beem:2013sza} are correct.} and assuming the existence of a GVSI, $\varphi_{n,N}$, described above, we have the following bound on the number of strong generators of $\chi[(n,N)]$ relative to the number of strong generators of $\chi[\mathfrak{su}(N)]$ 
\begin{equation}\label{nNN4}
|\chi[(n,N)]|\le|\chi[\mathfrak{su}(N)]|-4~.
\end{equation}
Moreover, $\CA(6)$ is the unique theory saturating \eqref{nNN4} in the class of $(n,N)$ SCFTs.

\noindent
{\bf Proof:} Let us first consider the case of $n=2$. We require that ${\rm gcd}(N,n)=1$ and therefore that $N\ge3$. From \eqref{hrelnN}, we see that the $SU(2)_F$ generators of the SYM theory are mapped to dimension two operators: $T$, $X_1$, and $X_2$. Here $T$ is the energy-momentum tensor and $X_{1,2}$ are other spin-two currents in $\chi[(2,N)]$. These must be strong generators since the $(2,N)$ theory has no flavor symmetries. By the conjecture for $\chi[\mathfrak{su}(N)]$ in \cite{Beem:2013sza}, there are no other bosonic generators at $h_{(2,N)}=2$.

At $h_{(2,N)}=3$ we will have the bosonic operators $\partial T$, $\partial X_1$, and $\partial X_2$. There cannot be any relation involving just these operators (otherwise some linear combination of $T$, $X_1$, and $X_2$ would be constant), and we also see from \eqref{hrelnN} that $\varphi_{2,N}(\partial J^-)$, $\varphi_{2,N}(\partial J^0)$, $\varphi_{2,N}(\partial J^+)$ can only contribute at $h_{(2,N)}=4$. Therefore, we learn that three strong generators of $\chi[\mathfrak{su}(N)]$ must map to the $h_{(2,N)}=3$ bosonic derivatives in $\chi[(2,N)]$. These are the 2D states arising from three of the four $B_{ijk}$ operators of the $\mathfrak{su}(N)$ $\CN=4$ SYM VOA conjectured in \cite{Beem:2013sza} (note that here we are using the fact that $N\ge3$).

At $h_{(2,N)}=3$ we also have four fermionic states arising from mapping $\left\{G^{\pm},\tilde G^{\pm}\right\}$ to $\left\{\Psi^{\pm},\tilde\Psi^{\pm}\right\}$ under $\varphi_{2,N}$. By similar logic to the one used above in the bosonic case, we must have four independent derivatives at $h_{(2,N)}=4$: $\partial\Psi^{\pm}$ and $\partial\tilde\Psi^{\pm}$. From \eqref{hrelnN} we see that these states cannot arise from images of $\varphi(\partial G^{\pm})$ or $\varphi(\partial\tilde G^{\pm})$. Therefore, these must be generated by four fermionic currents in $\chi[\mathfrak{su}(N)]$ (these are four of the 2D avatars of the 4D $\Tr Q_iQ_j\tilde\lambda_{\dot+}$ and $\Tr Q_iQ_j\lambda_{\dot+}$ operators). As a result, we find the stronger result that
\begin{equation}
|\chi[(n,N)]|\le|\chi[(\mathfrak{su}(N)]|-7<\chi[\mathfrak{su}(N)]|-4~.
\end{equation}

Consider now the case of $n=3$. Just as in the discussion of the $(3,2)$ theory in the previous sections, we see that \eqref{hrelnN} implies that the first non-trivial operator of $\chi[(3,N)]$ enters at $h_{(3,N)}=2$: the stress tensor, $T$. We again have the unique identification, $\varphi(J^-)=T$. At $h_{(3,N)}=3$ we have $\partial T\ne0$. Since \eqref{hrelnN} implies $\varphi(\partial J^{-})$ has $h_{(3,N)}=5$, we see that a strong generator of $\chi[\mathfrak{su}(N)]$ must map to $\partial T$. By the conjecture in \cite{Beem:2013sza} we have that 
\begin{equation}
\varphi(\alpha\cdot b_{111}+\beta\cdot J^0)=\partial T\in\chi[(2,N)]~,
\end{equation}
for some constants $\alpha$ and $\beta$. Here $b_{111}=\chi(B_{111})$ where $B_{111}=\Tr Q_1Q_1Q_1$ (\lq\lq$1$" is a label denoting the lowest $SU(2)_F$ weight state in the doublet $Q_i$). An independent linear combination of $\varphi(b_{111})$ and $\varphi(J^0)$ map to a $\chi[(3,N)]$ strong generator, $X$, at $h_{(3,N)}=3$ (this state was absent in the case $N=2$, but it exists for $N\ge4$).

Next consider the bosonic states at $h_{(3,N)}=4$. We have $T^2$, $\partial^2T$, and $\partial X$. We can't have a null state involving just $\partial^2T$ and $\partial X$ (since this would contradict $\partial T$ and $X$ being independent operators at $h_{(3,N)}=3$) \footnote{Moreover, since $c\ne-{22\over5}$, we also cannot have a null relation involving just $\partial^2T$ and $T^2$.}. Therefore, we have that $\partial^2T$ and $\partial X$ are independent. Since $\varphi(\partial b_{111})$ has $h_{(3,N)}=6$, we see that two strong generators of $\chi[\mathfrak{su}(N)]$ must map to the derivatives in question. This mapping occurs via two linear combinations of $\varphi(J^+)$, $\varphi(b_{(211)})$ (where the indices of $b_{(211)}$ are symmetrized), and $\varphi(b_{1111})$. We therefore see that there is an additional independent generator at this level, $Y$ (again only for $N\ge4$).

To finish off the case of $n=3$, let us consider the fermionic states at $h_{(3,N)}=4$. We have two states arising from the mapping of $\left\{G^{-},\tilde G^{-}\right\}$ to $\left\{\Psi,\tilde\Psi\right\}$ under $\varphi_{3,N}$. Therefore, at $h_{(3,N)}=5$, we have $\partial\Psi$ and $\partial\tilde\Psi$. By \eqref{hrel}, these cannot arise from $\varphi(\partial G^-)$ or $\varphi(\partial \tilde G^-)$. As a result, two strong generators of $\chi[\mathfrak{su}(N)]$ must map onto them. We conclude that
\begin{equation}
|\chi[(3,N)]|\le|\chi[\mathfrak{su}(N)]|-4~,
\end{equation}
with the equality saturated if and only if $N=2$.

Consider now the case of $n=4$ (since ${\rm gcd}(4,N)=1$, we see that $N\ge3$). Again our first bosonic operator in $\chi[(4,N)]$ enters at $h_{(4,N)}=2$ and has a unique mapping: $\varphi(J^-)=T$. As in the previous case, at $h_{(4,N)}=3$ we have $\partial T\ne0$. Since \eqref{hrelnN} implies that $\varphi(\partial J^-)$ has $h_{(4,N)}=6$, we see that a strong generator of $\chi[\mathfrak{su}(N)]$ must map to $\partial T$. By the conjecture in \cite{Beem:2013sza}, we have that $\varphi(b_{111})=\partial T$. Next, at $h_{(4,N)}=4$, we have $T^2$ and $\partial^2T\ne0$ (since $c\ne-{22\over5}$ we also have $\partial^2T$ is independent of $T^2$). Since $\varphi(\partial b_{111})$ has $h_{(4,N)}=7$, we must have $\varphi(J^0)$, $\varphi(J^-J^-)$, and $\varphi(b_{1111})$ mapping into these stress tensor states and a new strong generator $X$ (for $N\ge5$; we see that for $N=3$, $X$ cannot exist using \cite{Beem:2013sza}). Since $\varphi(\partial J^0)$, $\varphi(\partial J^-J^-)$,  $\varphi(\partial b_{1111})$ have $h_{(4,N)}=8$, this means there is a strong generator that maps to $\partial^3T$.

Now consider the lowest-dimensional fermions $\varphi(G^-)=\Psi$ and $\varphi(\tilde G^-)=\tilde\Psi$ at $h_{(4,N)}=5$. These are strong generators of $\chi[(4,N)]$, and their derivatives $\partial\Psi$ and $\partial\tilde\Psi$ must correspond to other strong generators of the $\chi[\mathfrak{su}(N)]$ VOA by \eqref{hrelnN}. Therefore, we have the stronger result that 
\begin{equation}
|\chi[(4,N)]|\le|\chi[\mathfrak{su}(N)]|-5<|\chi[\mathfrak{su}(N)]|-4~.
\end{equation}

Finally, consider the case of $n=6$ (since ${\rm gcd}(6,N)=1$, we see that $N\ge5$). As above, our first bosonic operator in $\chi[(6,N)]$ enters at $h_{(6,N)}=2$ and has a unique mapping: $\varphi(J^-)=T$. Again at $h_{(6,N)}=3$ we have $\partial T\ne0$. Since \eqref{hrelnN} implies that $\varphi(\partial J^-)$ has $h_{(6,N)}=8$, we see that a strong generator of $\chi[\mathfrak{su}(N)]$ must map to $\partial T$. By the conjecture in \cite{Beem:2013sza}, we have that $\varphi(b_{111})=\partial T$. Next, at $h_{(8,N)}=4$ we have $T^2$ and $\partial^2T\ne0$ (again, since $c\ne-{22\over5}$, we also have that $\partial^2T$ is independent of $T^2$). Since $\varphi(\partial b_{111})$ has $h_{(6,N)}=9$, we must have $\varphi(J^-J^-)$ and $\varphi(b_{1111})$ mapping into linear combinations of $T^2$ and $\partial^2T$. Since $\varphi(\partial b_{1111})$ has $h_{(6,N)}=10$, we must have another strong generator mapping into $\partial^3T$

Now consider the lowest-dimensional fermions $\varphi(G^-)=\Psi$ and $\varphi(\tilde G^-)=\tilde\Psi$ at $h_{(6,N)}=7$. These are strong generators of $\chi[(6,N)]$, and their derivatives, $\partial\Psi$ and $\partial\tilde\Psi$, must correspond to other strong generators of the $\chi[\mathfrak{su}(N)]$ VOA by \eqref{hrelnN}.Therefore, we have the stronger result that 
\begin{equation}
|\chi[(6,N)]|\le|\chi[\mathfrak{su}(N)]|-5<|\chi[\mathfrak{su}(N)]|-4~,
\end{equation}
and we have proved our result. $\square$

\section{conclusions}\label{conclusions}
In this paper, we have shown that a surprisingly simple index relation between two very different theories---the $(3,2)$ AD theory and $\mathfrak{su}(2)$ $\CN=4$ SYM---has a mathematical explanation in terms of an exact GVSI. We've argued that physically this relation is a (distant) cousin of spontaneous symmetry breaking. It would be interesting to understand this physical perspective better, perhaps making contact with and generalizing the large charge literature as in \cite{Hellerman:2017sur,Hellerman:2017veg}.

It will also be interesting to understand how general the above phenomena are. As we have teased, we will find infinitely many generalizations of some of the above arguments in an upcoming work \cite{2020} (though note that the index discussion from the introduction referencing \cite{Gaiotto:2012xa} does not directly apply to the case $n\ne3$).

Finally, we have seen that the VOA associated with the $(3,2)$ SCFT, $\CA(6)$, saturates two bounds on the number of strong generators. It would be interesting to understand if this saturation is related to the fact that the $(3,2)$ theory lacks a Higgs branch. Moreover, the saturation of these bounds might imply that this theory can be targeted in interesting ways with the bootstrap.

\acknowledgements{We thank H.~Jiang for discussions and J.~Shafiq for collaboration on the early stages of this project. M.~B.'s work is supported by the Royal Society under the grants \lq\lq New Constraints and Phenomena in Quantum Field Theory" and \lq\lq New Aspects of Conformal and Topological Field Theories Across Dimensions." T.~N.'s work is supported by
JSPS Grant-in-Aid for Early-Career Scientists 18K13547.}


\begin{thebibliography}{}

%\cite{Buican:2016arp}
\bibitem{Buican:2016arp}
M.~Buican and T.~Nishinaka,
``Conformal Manifolds in Four Dimensions and Chiral Algebras,''
J. Phys. A \textbf{49}, no.46, 465401 (2016)
%doi:10.1088/1751-8113/49/46/465401
[arXiv:1603.00887 [hep-th]].
%32 citations counted in INSPIRE as of 04 Dec 2020

%\cite{DelZotto:2015rca}
\bibitem{DelZotto:2015rca}
M.~Del Zotto, C.~Vafa and D.~Xie,
``Geometric engineering, mirror symmetry and $ 6{\mathrm{d}}_{\left(1,0\right)}\to 4{\mathrm{d}}_{\left(\mathcal{N}=2\right)} $,''
JHEP \textbf{11}, 123 (2015)
%doi:10.1007/JHEP11(2015)123
[arXiv:1504.08348 [hep-th]].

%\cite{Argyres:1995xn}
\bibitem{Argyres:1995xn}
P.~C.~Argyres, M.~R.~Plesser, N.~Seiberg and E.~Witten,
``New N=2 superconformal field theories in four-dimensions,''
Nucl. Phys. B \textbf{461}, 71-84 (1996)
%doi:10.1016/0550-3213(95)00671-0
[arXiv:hep-th/9511154 [hep-th]].

%\cite{Cecotti:2012jx}
\bibitem{Cecotti:2012jx}
S.~Cecotti and M.~Del Zotto,
``Infinitely many N=2 SCFT with ADE flavor symmetry,''
JHEP \textbf{01}, 191 (2013)
%doi:10.1007/JHEP01(2013)191
[arXiv:1210.2886 [hep-th]].
%40 citations counted in INSPIRE as of 04 Dec 2020

%\cite{Cecotti:2013lda}
\bibitem{Cecotti:2013lda}
S.~Cecotti, M.~Del Zotto and S.~Giacomelli,
``More on the N=2 superconformal systems of type $D_p(G)$,''
JHEP \textbf{04}, 153 (2013)
%doi:10.1007/JHEP04(2013)153
[arXiv:1303.3149 [hep-th]].
%43 citations counted in INSPIRE as of 04 Dec 2020

%\cite{Cecotti:2010fi}
\bibitem{Cecotti:2010fi}
S.~Cecotti, A.~Neitzke and C.~Vafa,
``R-Twisting and 4d/2d Correspondences,''
[arXiv:1006.3435 [hep-th]].
%200 citations counted in INSPIRE as of 04 Dec 2020

\bibitem{Beem:2013sza}
C.~Beem, M.~Lemos, P.~Liendo, W.~Peelaers, L.~Rastelli and B.~C.~van Rees,
``Infinite Chiral Symmetry in Four Dimensions,''
Commun. Math. Phys. \textbf{336}, no.3, 1359-1433 (2015)
%doi:10.1007/s00220-014-2272-x
[arXiv:1312.5344 [hep-th]].

%\cite{Buican:2015ina}
\bibitem{Buican:2015ina}
M.~Buican and T.~Nishinaka,
``On the superconformal index of Argyres\textendash{}Douglas theories,''
J. Phys. A \textbf{49}, no.1, 015401 (2016)
%doi:10.1088/1751-8113/49/1/015401
[arXiv:1505.05884 [hep-th]].
%92 citations counted in INSPIRE as of 04 Dec 2020

%\cite{Buican:2015hsa}
\bibitem{Buican:2015hsa}
M.~Buican and T.~Nishinaka,
``Argyres\textendash{}Douglas theories, S$^1$ reductions, and topological symmetries,''
J. Phys. A \textbf{49}, no.4, 045401 (2016)
%doi:10.1088/1751-8113/49/4/045401
[arXiv:1505.06205 [hep-th]].
%33 citations counted in INSPIRE as of 04 Dec 2020

%\cite{Cordova:2015nma}
\bibitem{Cordova:2015nma}
C.~Cordova and S.~H.~Shao,
``Schur Indices, BPS Particles, and Argyres-Douglas Theories,''
JHEP \textbf{01}, 040 (2016)
%doi:10.1007/JHEP01(2016)040
[arXiv:1506.00265 [hep-th]].
%105 citations counted in INSPIRE as of 04 Dec 2020

%\cite{Buican:2015tda}
\bibitem{Buican:2015tda}
M.~Buican and T.~Nishinaka,
``Argyres-Douglas Theories, the Macdonald Index, and an RG Inequality,''
JHEP \textbf{02}, 159 (2016)
%doi:10.1007/JHEP02(2016)159
[arXiv:1509.05402 [hep-th]].
%42 citations counted in INSPIRE as of 04 Dec 2020

%\cite{Feigin:2007sp}
\bibitem{Feigin:2007sp}
B.~Feigin, E.~Feigin and I.~Tipunin,
``Fermionic formulas for (1,p) logarithmic model characters in Phi\{2,1\} quasiparticle realisation,''
[arXiv:0704.2464 [hep-th]].

%\cite{Feigin:2008sg}
\bibitem{Feigin:2008sg}
B.~L.~Feigin and I.~Y.~Tipunin,
``Characters of coinvariants in (1,p) logarithmic models,''
[arXiv:0805.4096 [math.QA]].

%\cite{Gaiotto:2012xa}
\bibitem{Gaiotto:2012xa}
D.~Gaiotto, L.~Rastelli and S.~S.~Razamat,
``Bootstrapping the superconformal index with surface defects,''
JHEP \textbf{01}, 022 (2013)
%doi:10.1007/JHEP01(2013)022
[arXiv:1207.3577 [hep-th]].
%173 citations counted in INSPIRE as of 04 Dec 2020

%\cite{Gadde:2011uv}
\bibitem{Gadde:2011uv}
A.~Gadde, L.~Rastelli, S.~S.~Razamat and W.~Yan,
``Gauge Theories and Macdonald Polynomials,''
Commun. Math. Phys. \textbf{319}, 147-193 (2013)
%doi:10.1007/s00220-012-1607-8
[arXiv:1110.3740 [hep-th]].

%\cite{Xie:2016evu}
\bibitem{Xie:2016evu}
D.~Xie, W.~Yan and S.~T.~Yau,
``Chiral algebra of Argyres-Douglas theory from M5 brane,''
[arXiv:1604.02155 [hep-th]].
%46 citations counted in INSPIRE as of 04 Dec 2020

%\cite{kac2017remark}
\bibitem{kac2017remark}
V.~G.~Kac and M.~Wakimoto,
``A remark on boundary level admissible representations''
Comptes Rendus Mathematique {\bf355}, no.2 128 (2017)
[arXiv:1612.07423  [math.RT]].
%46 citations counted in INSPIRE as of 04 Dec 2020

%\cite{Creutzig:2017qyf}
\bibitem{Creutzig:2017qyf}
T.~Creutzig,
``W-algebras for Argyres-Douglas theories,''
[arXiv:1701.05926 [hep-th]].

%\cite{Buican:2017fiq}
\bibitem{Buican:2017fiq}
M.~Buican, Z.~Laczko and T.~Nishinaka,
``$ \mathcal{N} $ = 2 S-duality revisited,''
JHEP \textbf{09}, 087 (2017)
%doi:10.1007/JHEP09(2017)087
[arXiv:1706.03797 [hep-th]].

%\cite{Buican:2017rya}
\bibitem{Buican:2017rya}
M.~Buican and Z.~Laczko,
``Nonunitary Lagrangians and unitary non-Lagrangian conformal field theories,''
Phys. Rev. Lett. \textbf{120}, no.8, 081601 (2018)
%doi:10.1103/PhysRevLett.120.081601
[arXiv:1711.09949 [hep-th]].

%\cite{Buican:2019huq}
\bibitem{Buican:2019huq}
M.~Buican and Z.~Laczko,
%``Rationalizing CFTs and Anyonic Imprints on Higgs Branches,''
JHEP \textbf{03}, 025 (2019)
doi:10.1007/JHEP03(2019)025
[arXiv:1901.07591 [hep-th]].

%\cite{Buican:2019kba}
\bibitem{Buican:2019kba}
M.~Buican, L.~Li and T.~Nishinaka,
``Peculiar Index Relations, 2D TQFT, and Universality of SUSY Enhancement,''
JHEP \textbf{01}, 187 (2020)
%doi:10.1007/JHEP01(2020)187
[arXiv:1907.01579 [hep-th]].
%1 citations counted in INSPIRE as of 04 Dec 2020

%\cite{2020}
\bibitem{2020}
{\it To appear}.

%\cite{Buican:2013ica}
\bibitem{Buican:2013ica}
M.~Buican,
``Minimal Distances Between SCFTs,''
JHEP \textbf{01}, 155 (2014)
%doi:10.1007/JHEP01(2014)155
[arXiv:1311.1276 [hep-th]].

%\cite{Bonetti:2018fqz}
\bibitem{Bonetti:2018fqz}
F.~Bonetti, C.~Meneghelli and L.~Rastelli,
``VOAs labelled by complex reflection groups and 4d SCFTs,''
JHEP \textbf{05}, 155 (2019)
%doi:10.1007/JHEP05(2019)155
[arXiv:1810.03612 [hep-th]].

%\cite{Dolan:2002zh}
\bibitem{Dolan:2002zh}
F.~A.~Dolan and H.~Osborn,
``On short and semi-short representations for four-dimensional superconformal symmetry,''
Annals Phys. \textbf{307}, 41-89 (2003)
%doi:10.1016/S0003-4916(03)00074-5
[arXiv:hep-th/0209056 [hep-th]].

%\cite{Dobrev:1985qv}
\bibitem{Dobrev:1985qv}
V.~K.~Dobrev and V.~B.~Petkova,
``All Positive Energy Unitary Irreducible Representations of Extended Conformal Supersymmetry,''
Phys. Lett. B \textbf{162}, 127-132 (1985)
%doi:10.1016/0370-2693(85)91073-1

%\cite{Creutzig:2018ltv}
\bibitem{Creutzig:2018ltv}
T.~Creutzig, D.~Gaiotto and A.~R.~Linshaw,
``S-duality for the Large $N = 4$ Superconformal Algebra,''
Commun. Math. Phys. \textbf{374}, no.3, 1787-1808 (2020)
%doi:10.1007/s00220-019-03673-4
[arXiv:1804.09821 [math.QA]].

%\cite{Hellerman:2017sur}
\bibitem{Hellerman:2017sur}
S.~Hellerman and S.~Maeda,
``On the Large $R$-charge Expansion in ${\mathcal N} = 2$ Superconformal Field Theories,''
JHEP \textbf{12}, 135 (2017)
%doi:10.1007/JHEP12(2017)135
[arXiv:1710.07336 [hep-th]].

%\cite{Hellerman:2017veg}
\bibitem{Hellerman:2017veg}
S.~Hellerman, S.~Maeda and M.~Watanabe,
%``Operator Dimensions from Moduli,''
JHEP \textbf{10}, 089 (2017)
%doi:10.1007/JHEP10(2017)089
[arXiv:1706.05743 [hep-th]].

\end{thebibliography}
\end{document}